%% file: main.tex
\documentclass[5p,times]{elsarticle}

\usepackage{amsmath,amssymb,amsfonts}
\usepackage{algorithmic}
\usepackage{algorithm}
\usepackage{graphicx}
\usepackage{xcolor}
\usepackage{colortbl}
\usepackage{color, soul}
\usepackage{lineno,hyperref}
\usepackage{booktabs}
\usepackage{multirow}
\usepackage{subfig}
\usepackage[skip=1mm]{caption}
\usepackage{subfig}
\usepackage{microtype}
\modulolinenumbers[5]

\journal{Parallel Computing}

\begin{document}

\begin{frontmatter}

\title{Sphynx: a parallel multi-GPU graph partitioner for distributed-memory systems}

\author{Seher Acer\fnref{myfootnote}\corref{mycorrespondingauthor}}
\author{Erik G. Boman\fnref{myfootnote}\corref{mycorrespondingauthor}}
\author{Christian A. Glusa\fnref{myfootnote}}
\author{Sivasankaran Rajamanickam\fnref{myfootnote}}
\address{Center for Computing Research, Sandia National Laboratories, Albuquerque, NM, U.S.A.}
\fntext[myfootnote]{Email addresses: sacer@sandia.gov (S. Acer), egboman@sandia.gov (E. G. Boman), caglusa@sandia.gov (C. A. Glusa), and srajama@sandia.gov (S. Rajamanickam)}
\cortext[mycorrespondingauthor]{Corresponding author}

\begin{abstract}

Graph partitioning has been an important tool to partition the work among 
several processors to minimize the communication cost and balance the workload.
While accelerator-based supercomputers are emerging to be the standard, the use of
graph partitioning becomes even more important as applications are rapidly moving to 
these architectures. However, there is no distributed-memory-parallel,
multi-GPU graph partitioner available for applications. We developed a spectral
graph partitioner, Sphynx, using the portable, accelerator-friendly stack of the Trilinos
framework. In Sphynx, we allow using different preconditioners and exploit their unique advantages.
We use Sphynx to systematically evaluate the various algorithmic choices
in spectral partitioning with a focus on the GPU performance. We perform those evaluations
on two distinct classes of graphs: \emph{regular} (such as meshes, matrices from finite element methods) and 
\emph{irregular} (such as social networks and web graphs), and show that different
settings and preconditioners are needed for these graph classes.
The experimental results on the Summit supercomputer show that Sphynx is the fastest alternative on irregular graphs in an application-friendly setting and obtains a partitioning quality close to ParMETIS on regular graphs.
When compared to nvGRAPH on a single GPU, Sphynx is faster and obtains better balance and better quality partitions.  
Sphynx provides a good and robust partitioning method across a wide range of graphs for applications looking for a GPU-based partitioner. 

\end{abstract}

\begin{keyword}
graph partitioning, spectral partitioning, GPUs, distributed-memory systems.
\end{keyword}

\end{frontmatter}


\input{introduction}
\input{related}
\input{background}
\input{sphynx}

\input{preconditioners}

\input{experiments}
\input{conclusions}

\section*{Acknowledgments}
We thank Karen Devine and Kamesh Madduri for the helpful discussions.
We thank Jennifer Loe for guidance on using the polynomial preconditioner.
Sandia National Laboratories is a multimission laboratory managed and operated by National Technology and Engineering Solutions of Sandia, LLC., a wholly owned subsidiary of Honeywell International, Inc., for the U.S. Department of Energy's National Nuclear Security Administration under contract DE-NA-0003525.
This research was supported by the Exascale Computing Project (17-SC-20-SC), a collaborative effort of the U.S. Department of Energy Office of Science and the National Nuclear Security Administration.
This paper describes objective technical results and analysis. Any subjective views or opinions that might be expressed in the paper do not necessarily represent the views of the U.S. Department of Energy or the United States Government.

\bibliography{references}

\end{document}

%% file: introduction.tex
\section{Introduction}
\label{sec:intro}
Partitioning data for large parallel simulations from scientific computing and
distributed data analysis has been studied for decades. Partitioning methods 
vary in the representation of the data (graphs, hypergraphs, coordinates), the cost of
the method, the metric they try to optimize, and the constraints in the problem.
Graph partitioning, hypergraph partitioning and spatial partitioning of a domain are all
popular methods~\cite{Metis,Catalyurek1999,Deveci2016}.
Graph partitioning has been particularly popular because of the software availability, its
performance (as opposed to hypergraph partitioning methods)
and easy representation of different types of data as graphs. Despite its popularity, no existing graph partitioner can effectively leverage accelerator-based distributed-memory architectures. The last three large supercomputers announced at U.S. DOE 
facilities are all accelerator-based platforms. The lack of a scalable graph partitioner on
accelerator-based architectures is a
key deficiency that parallel computing community has yet to address. Furthermore, 
the DOE facilities have all announced different accelerator hardware (AMD, Intel, and NVIDIA).
This adds another dimension to the problem, where the need is for a scalable, distributed-memory, accelerator-based
partitioner that is also portable to the different accelerators coming in the next few years.

The most important graph partitioning algorithm is the multilevel method \cite{Hendrickson95},
implemented in popular partitioning software such as METIS/ParMETIS \cite{Metis,ParMetis} and Scotch \cite{Scotch}.
However, this algorithm is difficult to parallelize on distributed-memory systems
as it relies on methods that are inherently sequential and have highly irregular memory access patterns. 
The parallelization becomes even harder on current accelerator hardware.
However, this is exactly what is needed by the applications that are running on the accelerator
hardware and the data is resident on the accelerator. An alternative is to move the data to the CPU and
call a CPU-based partitioner. However, the cost of moving data to CPU and back is prohibitive,
especially when applications require dynamic repartitioning of the data as the simulation progresses.

We revisit the graph partitioning problem on accelerators using a different approach - spectral partitioning.
Spectral methods have been used for both partitioning \cite{PothenSimonLiou90} and clustering 
\cite{ShiMalik00}, however some of the work is quite old
and many variations have been proposed. Multilevel partitioning methods became more popular
than spectral methods likely due to slight advantages in runtime and quality. 
We hypothesize some of it is due to the availability of software and evaluation
on problem sizes and architectures of interest two decades ago.
We revisit the spectral approach as we believe it has several unique advantages to the current problem 
and current architectures.
One major advantage
of spectral partitioning is that it is completely based on linear-algebra operations that are well-suited 
for GPUs.
It is based on eigenvectors of the graph Laplacian. A key advantage of this method
is that the main computation is now a sparse eigensolver,
so the main kernels are SpMV, dense matrix/vector operations and an optional preconditioner apply.
Thus, we can leverage decades of work in the HPC community on linear-algebra operations.
Our new partitioner, Sphynx, is based on Trilinos~\cite{trilinos-website} 
and uses Kokkos~\cite{Edwards2014} for portability across CPUs and GPUs.

There are a number of algorithmic and implementation choices for spectral partitioning
that have been developed over the years. However, they have not been evaluated with the 
goal of distributed-memory partitioning. 
We identify the key algorithmic choices and evaluate them using Sphynx to
determine what works best for distributed-memory spectral partitioning on GPUs.
These choices include which eigenvalue problem to solve, which eigensolver convergence 
tolerance to use, and which preconditioner to use.
We perform our evaluations on both regular and irregular graphs.


A preliminary version of this paper~\cite{AcerBomanRajamanickam-Sphynx} only focused on the performance on irregular graphs. The main contributions of ~\cite{AcerBomanRajamanickam-Sphynx}  can be summarized as follows:
\begin{itemize}
	\item We developed a portable, multi-GPU, distributed-memory
	parallel, spectral graph partitioner called Sphynx. To the best
	of our knowledge, this is the first graph partitioner for multi-GPU systems.
	\item We conducted a systematic study of several algorithmic choices for spectral partitioning 
	within the context of GPU-based partitioner performance on irregular graphs. 
	\item We identified performance problems due to the key kernels in the eigensolver,
	improved their performance,
	and demonstrate speedup of up to 14.8x compared to NVIDIA's cuBLAS.
	\item We evaluated Sphynx's performance against the state-of-the-art partitioner ParMETIS~\cite{ParMetis}.
\end{itemize}

This paper extends~\cite{AcerBomanRajamanickam-Sphynx} in two major directions: different graph types and preconditioners. The contributions of this paper can be summarized as follows:  
\begin{itemize}
	\item We evaluate Sphynx's performance with Jacobi, polynomial, and MueLu preconditioners
	in the LOBPCG eigensolver and incorporate them in the systematic study of algorithmic choices.
	This evaluation also shows the flexibility of Sphynx, which allows easy use of new preconditioners.  
	\item We evaluate Sphynx's performance on both regular graphs
        (such as meshes, matrices from finite element methods), and highly irregular graphs
	such as web graphs and social networks and demonstrate its performance benefits
	on GPUs. 
	\item We provide a framework for Sphynx with the optimal parameter settings customized for different 
	graph types and preconditioners.
	\item In addition to ParMETIS, we include XtraPuLP~\cite{xtrapulp} and nvGRAPH~\cite{Naumov16} in the experiments by which we compare Sphynx against the state of the art.
\end{itemize}

The experimental results on Summit show that Sphynx is the fastest alternative on irregular graphs in an application-friendly setting and obtains a partitioning quality close to ParMETIS on regular graphs.
When compared to nvGRAPH on a single GPU, Sphynx is faster and obtains better quality partitions.   
Sphynx has been released in Trilinos\footnote[1]{https://github.com/trilinos/Trilinos/} \cite{trilinos-website} as a subpackage of Zoltan2~\cite{Boman2012}.

The rest of the paper is organized as follows: Sections~\ref{sec:related} and~\ref{sec:background} cover the related work and background for spectral partitioning, respectively. Details of Sphynx and related preconditioners are described in Sections~\ref{sec:spbytrilinos} and \ref{sec:prec}, respectively. Finally, Section~\ref{sec:exp} gives the experimental results and Section~\ref{sec:conc} concludes the paper.

%% file: related.tex
\section{Related Work}
\label{sec:related}

Spectral partitioning was first proposed in the 1970s 
\cite{Fiedler73,Donath72} but this work was mostly theoretical. 
The first practical implementation with impact in scientific computing
was by Pothen et al. \cite{PothenSimonLiou90}. They also showed that
the spectral bisection is a continuous relaxation of the minimum balanced cut problem.
Spectral partitioning was an option in Chaco~\cite{chaco}. 
Early work focused on bisection, and recursive bisection was used
for general number of parts. The use of multiple eigenvectors
was investigated in \cite{Alpert95dac,Hendrickson95spectral}.

Spectral partitioning was later adapted to clustering \cite{ShiMalik00}, 
where the cut objective was the ratio or normalized cut. For this problem, 
the normalized graph Laplacian is often superior to the standard (combinatorial)
Laplacian. A parallel (distributed memory) spectral clustering code
was developed by Chen et al. \cite{ChenSong11}. 
A spectral partitioner (targeted clustering) for GPU was developed by 
Naumov \cite{Naumov16} and is available in the NVIDIA Graph Analytics (nvGRAPH) library. However, 
it is limited to a single GPU and only solves the clustering problem via
ratio or normalized cuts.

Popular graph partitioners that do not use a spectral approach include 
ParMETIS~\cite{ParMetis}, Scotch \cite{Scotch}, KaHip~\cite{sandersschulz2013}, 
ParHip~\cite{ParHip}, PuLP~\cite{pulp,pulpjrnl}, and XtraPuLP~\cite{xtrapulp}.
Partitioners that use a hypergraph-based approach include PaToH~\cite{Catalyurek1999}, 
PHG (Zoltan/Zoltan2) \cite{zoltan} and Mondriaan~\cite{VastenhouwBisselling05}.  
These only run on CPU-based systems, not on GPUs.

%% file: background.tex
\section{Background}
\label{sec:background}
\subsection{Graph Partitioning Problem}
\label{sec:gp}
An undirected graph $\cal G=(\cal V,\cal E)$ is defined as a set of vertices $\cal V$ and a set of edges $\cal E$, where each edge $e_{i,j}$ connects two distinct vertices $v_i$ and $v_j$. 
Each vertex $v_i$ is associated with a weight denoted by $w(v_i)$ and each edge $e_{i,j}$ is associated with a cost denoted by $c(e_{i,j})$.

$\Pi=\{{\cal V}_1,{\cal V}_2, \ldots, {\cal V}_K\}$ is a $K$-way partition of $\cal G$ if parts are non-empty and mutually exclusive and the union of the parts gives $\cal V$. 
An edge $e_{i,j}$ is said to be cut in $\Pi$ if vertices $v_i$ and $v_j$ belong to different parts, and uncut/internal, otherwise. Let ${\cal E}_C$ denote the set of cut edges in $\Pi$. 
The cutsize of $\Pi$ is defined as the total cost of the cut edges, that is,
\begin{equation}
cutsize = \sum_{e_{i,j}\in{\cal E}_C}c(e_{i,j}).
\label{eq:cutsize}
\end{equation}

Each part ${\cal V}_k$ has a weight, $W_k$, which is computed as the sum of the weights of the vertices in ${\cal V}_k$, that is, $W_k=\sum_{v_i\in{\cal V}_k}w(v_i)$.
$\Pi$ is said to be balanced if, for a given maximum allowable imbalance ratio $\epsilon$, each part weight $W_k$ satisfies $W_k < W_{avg}(1+\epsilon)$.
Here, $W_{avg}$ denotes the average part weight, that is, $W_{avg}=\sum_{k=1}^K W_k/K=\sum_{v_i\in\cal V} w(v_i)/K$.

For a given graph $\cal G$, and $K$ and $\epsilon$ values, the graph partitioning problem is defined as finding a balanced $K$-way partition $\Pi$ of $\cal G$ which minimizes the cutsize. 

\subsection{Spectral Graph Partitioning}
\label{sec:sp}
Spectral graph partitioning utilizes one or more eigenvectors of the Laplacian matrix associated with the given graph $\cal G$.
The are several variations of the Laplacian matrix: the combinatorial Laplacian $L_C$,
 the normalized Laplacian $L_N$, or the generalized Laplacian method.
They are all symmetric and positive semidefinite matrices of size $n\times n$, where $n$ denotes the number of vertices in $\cal G$. Thus, the eigenvalues are real and non-negative.

Let $A$ and $D$ denote the adjacency matrix of $\cal G$ and a diagonal matrix holding the degrees of the vertices in $\cal G$, respectively. 
The combinatorial Laplacian $L_C$ is defined as $L_C=D-A$, that is,
\begin{equation}
L_C(i,j)=
\begin{cases}
  deg(v_i) & \text{if }i=j,\\    
  -1 & \text{if }i\neq j \text{ and } a_{i,j}\neq 0,\\    
  0 & \text{if }i\neq j \text{ and } a_{i,j}=0.\\    
\end{cases}
\label{eq:lp}
\end{equation}
The normalized Laplacian $L_N$ is defined as $L_N=I-D^{-1/2}AD^{-1/2}$, that is, 
\begin{equation}
L_N(i,j)=
\begin{cases}
  1 & \text{if }i=j,\\    
  -\frac{1}{\sqrt{deg(v_i)deg(v_j})} & \text{if }i\neq j \text{ and } a_{i,j}\neq 0,\\    
  0 & \text{if }i\neq j \text{ and } a_{i,j}=0.\\    
\end{cases}
\label{eq:lpn}
\end{equation}
For the combinatorial Laplacian $L_C$, the graph edge cut corresponds exactly to
$\frac{1}{4} x^t L_C x$, where $x$ is a vector of discrete values $\pm 1$. 
The key observation is that this discrete, combinatorial problem can be 
relaxed by solving for continuous values, which then gives the symmetric 
eigenvalue problem.
Thus, the problem is to find the eigenvectors $x$ corresponding to the smallest eigenvalues $\lambda$ that satisfy $Lx=\lambda x$, where $L$ is either $L_N$ or $L_C$.
We ignore the trivial eigenvector corresponding to $\lambda=0$.
For the generalized Laplacian, we solve the generalized eigenvalue problem
$L_C x=\lambda D x$. This is closely related to the normalized Laplacian $L_N$ but
the eigenvectors will differ.

Then, $\cal G$ is partitioned by using a geometric approach on the eigenvector entries, which are effectively used as vertex coordinates. 
One can represent weighted graphs by simply setting the off-diagonal matrix entries to the negative edge weights, and adjusting the diagonal entries to be the sum of the incident edge weights.
In this paper, we only study unweighted graphs but the algorithms extend easily
to weighted graphs.

In the earliest recursive spectral bisectioning methods \cite{PothenSimonLiou90}, just one eigenvector $x_F$, i.e., the Fiedler vector, is used at each recursive step.
The entries of eigenvector $x_F$ are sorted in increasing order and the vertices that correspond to the first and second halves of the sorted entries are assigned to the first and second parts, respectively. 
Then vertex induced subgraphs are formed on these parts and the same bisectioning procedure is applied recursively by finding new eigenvectors on these new graphs. 
An improved version of the recursive spectral partitioning~\cite{Hendrickson95spectral} incorporates non-unit vertex weights and computes two (or more) parts based on these weights, but the algorithmic details are fairly complicated.
This work also extends the idea of using one eigenvector for bisectioning to using two or three eigenvectors for quadrisectioning or octasectioning at each recursive step.

\subsection{LOBPCG}

\begin{algorithm}[t]
\begin{algorithmic}[1]
\REQUIRE {$A$, number of desired eigenvalues $nev$}
\vspace{0.5em}
\STATE Select an initial guess $\tilde{X} \in R^{n \times nev}$
\STATE $X_0 \gets \tilde{X}Y$, where $(Y, \Theta_0) \gets RR(\tilde{X},nev)$ \label{alg:RRinit}
\STATE $R_I \gets A X_0 - X_0 \Theta_0$ \label{alg:matvecinit}
\STATE $P_I \gets \left[ \right] $
\FOR{$k \gets 0,1,2, \ldots$}
    \STATE Solve $M H_I = R_I$ \emph{// Preconditioned linear system.} \label{alg:prec}
    \STATE $S \gets \left[ X_k, H_I, P_I \right]$ and compute $(Y, \Theta_{k+1}) = RR(S, nev)$ \label{alg:RR}
    \STATE $X_{k+1} \gets \left[ X_k, H_I, P_I \right]Y$ \label{alg:update}
    \STATE $R \gets A X_{k+1} - X_{k+1} \Theta_{k+1}$ \emph{// Residual} \label{alg:matvec}
    \STATE Set $R_I$ with the unconverged columns of $R$
    \STATE Set $Y_I$ with the columns of $Y$ associated with the unconverged columns of $R$
    \STATE $P_I \gets  \left[ 0, H_I, P_I \right] Y_I$
\ENDFOR
\vspace{0.5em}
\RETURN $X_k$
\end{algorithmic}
\caption{LOBPCG}
\label{alg:lobpcg}
\end{algorithm}

The LOBPCG algorithm \cite{Knyazev01} attempts to find extreme eigenpairs of
a matrix $A$ by minimizing (maximizing) the Rayleigh quotient
$x^T Ax/(x^T x)$. The key idea is to use Rayleigh-Ritz analysis
on a carefully chosen subspace. Specifically, this subspace
includes both the block of current iterates $X$, the previous
iterates, and the preconditioned residual. Allowing for a 
preconditioner is a key advantage over traditional methods like
Lanczos or Arnoldi.

A concise formulation of the LOBPCG algorithm was given
in \cite{Hetmaniuk06}. We restate in Algorithm~\ref{alg:lobpcg} for the 
simpler standard (not generalized) eigenvalue problem.
Here, $RR()$ denotes the Rayleigh-Ritz procedure, which is essentially
solving a smaller (projected) eigenproblem on a subspace.
We note the computationally expensive parts are (i) the block matrix-vector multiply (lines~\ref{alg:matvecinit} and \ref{alg:matvec}),
(ii) the RR step (lines~\ref{alg:RRinit} and \ref{alg:RR}), (iii) all the small dense linear algebra computations (lines~\ref{alg:RRinit}, \ref{alg:matvecinit}, \ref{alg:update} and \ref{alg:matvec}),
and potentially (iv) the preconditioning step (line~\ref{alg:prec}). We have previously
observed that for irregular graphs, the Jacobi preconditioner works
well as it captures the highly variable vertex degrees. Also, it
is computationally inexpensive and easy to do on GPU.

LOBPCG is widely used and implementations exist in Trilinos~\cite{trilinos-website}, Hypre~\cite{hypre},
and PETSc/SLEPc~\cite{petsc}. LOBPCG has been found to be particularly
effective for graph Laplacians \cite{BomanDevineLehoucq14}.

%% file: sphynx.tex
\section{SPHYNX: A Spectral Graph Partitioning Tool}
\label{sec:spbytrilinos}

In this section, we propose a new, parallel, spectral graph partitioning tool called Spyhnx.
The basic steps performed in Sphynx are given in Algorithm~\ref{alg:sp}.

\begin{algorithm}[!tbp]
\begin{algorithmic}[1]
\REQUIRE {$\cal G=(\cal V, \cal E)$, $weights$, desired number of parts $K$}
\STATE $L\gets$ createLaplacian$(\cal G)$
\STATE $d \gets \lfloor \log_2 K \rfloor +1$
\STATE {// Compute $d$ smallest eigenvectors of $L_N$}
\STATE $E \gets LOBPCG(L,d)$ 
\STATE {// Remove the first eigenvector}
\FOR{$i \gets 1$ \textbf{to} $V$}
  \FOR{$j \gets 1$ \textbf{to} $d-1$}
    \STATE $coords[i,j] \gets E[i,j+1]$ 
  \ENDFOR 
\ENDFOR
\STATE {// Compute $K$ vertex parts using $coords$ and $weights$}
\STATE $\Pi=\{{\cal V}_1,\ldots,{\cal V}_K\} \gets MJ(coords, weights, K)$ 
\RETURN $\Pi$
\end{algorithmic}
\caption{Sphynx}
\label{alg:sp}
\end{algorithm}

We first form the Laplacian $L$ of the input graph $\cal G=(\cal V, \cal E)$.
$L$ is either the combinatorial Laplacian $L_C$ or the normalized Laplacian $L_N$, depending on the problem type we use.
Recall that $L_C$ is used in both combinatorial and generalized eigenvalue problems.
Then we call the LOBPCG algorithm to obtain the eigenvectors that correspond to the $d$ smallest eigenvalues of $L_N$.
In lines 4 and 8 of Algorithm~\ref{alg:sp}, matrix $E$ denotes the set of returned eigenvectors, so, the $j^{th}$ column of $E$ corresponds to the $j^{th}$ smallest eigenvalue. 

We set the number of requested eigenvectors $d$ as 
\begin{equation}
d=\lfloor \log_2K \rfloor+1.
\label{eq:d}
\end{equation}
Sphynx computes $d$ eigenvectors, because unlike the traditional spectral graph partitioning algorithms, it adopts a $K$-way partitioning approach rather than a recursive-partitioning approach.
That is, it computes all eigenvectors to be used for partitioning at once and only on the Laplacian matrix of the original input graph $\cal G$.
Computing all eigenvectors at once \textit{avoids} 
\begin{itemize}
\item forming subgraphs and/or their corresponding Laplacian matrices in each recursive bipartitioning step,
\item moving the subgraphs across different nodes/GPUs of the distributed-memory system, and
\item calling LOBPCG multiple times, on different graphs.
\end{itemize} 

\begin{figure*}[!thbp]
\centering
\includegraphics[width=7in]{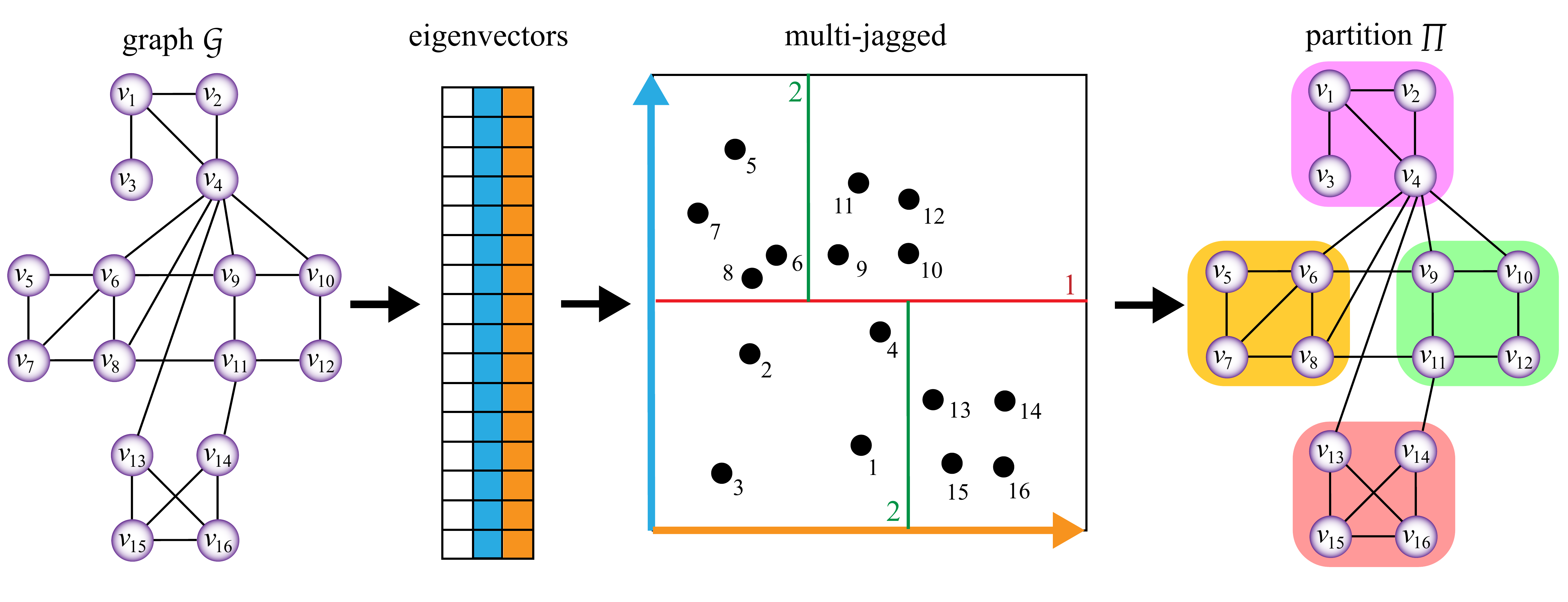}
\caption{Illustration of Sphynx partitioning a graph with 16 unit-weight vertices into 4 parts.}
\label{fig:example}
\end{figure*}

We use the eigenvectors stored in $E$ to embed the vertices of $\cal G$ into a low-dimensional coordinate space. 
While doing so, we exclude the first eigenvector, which corresponds to the smallest eigenvalue zero,
because it does not have useful embedding information as all its entries are one. 
Therefore, the coordinate space we embed our graph into is $(d-1)$-dimensional. 
Our embedding sets the $i^{th}$ entry of the $(j+1)^{st}$ eigenvector as the coordinate of vertex $v_i$ at the $j^{th}$ dimension, for $1 \leq i \leq V$ and $1 \leq  j < d$.

We use the multi-jagged (MJ) algorithm~\cite{Deveci2016} to partition the embedded graph.
MJ is a parallel geometric partitioning algorithm, which recursively obtains multi-sections on a given set of points (coordinates) along different dimensions.   
Note that this is different than sorting the entries of a single eigenvector, which was done by existing spectral partitioning algorithms to determine the bisection point.
The main objective of MJ is to maintain balance on the total weights of the sections/parts. 
MJ is quite flexible in the sense that it can partition any set of points in a given space into any number of parts, regardless of the number of dimensions of the space.
While doing so, it allows the user to specify the number of sections along each dimension. 
Another nice property of MJ is that during the recursive multi-sectioning, the new cut planes to be computed in a section do not need to confine to the existing cut planes in other sections.
This property provides more freedom in obtaining the balance on the part weights compared to a rectilinear partitioning approach. 
We use MJ with the default settings, which performs multisections along each dimension in a round-robin fashion.

Figure~\ref{fig:example} illustrates the steps performed by Sphynx on a graph with 16 vertices with unit weights.
In this example, the desired number of parts, $K$, is four, so, $d=\lfloor \log_24 \rfloor+1=3$ eigenvectors are computed.
 The entries of the second and the third eigenvectors are used as the coordinates of the vertices in the first and second dimensions, respectively.
MJ uses the coordinates to compute four parts by first computing the cut plane denoted by 1 and then computing the two cut planes denoted by 2.



Sphynx was implemented using the Trilinos framework~\cite{trilinos-website}.
Trilinos is a collection of object-oriented C++ packages implementing algorithms and technologies for the solution of large-scale, complex multi-physics engineering and scientific problems.
The packages used in Sphynx work well on distributed-memory architectures through message passing interface (MPI) for interprocessor communication. 

We use the Tpetra package~\cite{tpetra} to store and process the adjacency matrix $A$ of the input graph. 
Tpetra implements linear algebra objects in a hybrid-parallel manner by supporting MPI+X for various shared-memory parallel programming models including CUDA.
Our code takes the input graph in the adjacency matrix format stored in a \textit{Tpetra CrsMatrix} object.
Extension to \textit{Tpetra CrsGraph} as the input type is trivial.
Although CrsMatrix and CrsGraph both support any type of initial distribution, we use the default distribution provided by the default \textit{Tpetra Map} object.
In this distribution, each MPI process is assigned a block of consecutive rows, where each block has roughly the same number of rows.    

We create the Laplacian matrix $L$ as another Tpetra CrsMatrix object by reusing the existing sparsity structure (stored in a Tpetra CrsGraph object) of the input matrix $A$.
We create new arrays, \textit{Kokkos Views}, to hold the coefficients of the Laplacian and set their values by using the \textit{parallel\_for} functionality in Kokkos~\cite{Edwards2014}, which is an abstraction for parallel execution across different shared-memory parallel programming models.
Using Kokkos in both creating the arrays and setting their entries enables keeping and processing all the data needed by Sphynx in the GPUs. 

We use the LOBPCG implementation in the Anasazi package~\cite{Baker2009}, which implements several eigensolvers using Tpetra objects and functionality.
In LOBPCG, we use a block size equal to the number of eigenvectors requested, which is equal to $d=\lfloor \log_2K \rfloor+1$.

The MJ algorithm is provided by the Zoltan2 package~\cite{Boman2012}, which implements several partitioning, ordering, and coloring algorithms using Tpetra objects and functionality.

%% file: preconditioners.tex
\section{Preconditioners in Sphynx}
\label{sec:prec}

Preconditioning is an important technique to accelerate iterative methods. Traditionally, they have been used for linear systems $Ax=b$.
A preconditioner $M$ approximates $A$ in the sense that the preconditioned system $MA$ converges faster. It should also be inexpensive to compute and apply.
For eigensolvers, the picture is more complicated. Traditional eigensolvers like
Lanczos or Krylov-Schur cannot be preconditioned. However, the LOBPCG method can use
a preconditioner, and the same preconditioner for solving linear systems also 
improves convergence for LOBPCG. Therefore, we focus on preconditioners that
were developed for linear systems but use them in LOBPCG. 

There is a complex trade-off in preconditioning. The better quality of the
preconditioner, the fewer iterations are needed, but the more time is spent on the
preconditioner setup and apply. The best preconditioner for a particular problem class
is typically found empirically. Here we consider three popular preconditioners.

\subsection{Jacobi}
The Jacobi method is very simple: $M = diag(A)^{-1}$. Since the preconditioner is diagonal,
it is both cheap to compute and to apply. It is also well suited to GPUs.
The Jacobi method often works well on matrices from highly irregular graphs, as
the diagonal (vertex degrees) is important \cite{BomanDG16}. We use the Jacobi implementation
provided by the Ifpack2 package~\cite{ifpack2} in Trilinos.

\subsection{Polynomial Preconditioner}
We say $M = p_k(A)$ is a polynomial preconditioner if $p_k$ is a polynomial of degree $k$.
Given $p_k$, the preconditioner only requires SpMV to apply, so it is highly
parallel and well suited to GPUs. The challenge is to find a good polynomial.
Chebyshev polynomials require good estimates of the extreme eigenvalues. We use
a recent method that is more general, based on the GMRES polynomial \cite{Loe20,LoeArxiv}. We use the default value of 25 for the degree of the polynomial 
preconditioner. 
We use the implementation provided by the Belos package~\cite{belos} in Trilinos.

\input{muelu}

%% file: muelu.tex
\subsection{Multigrid (MueLu) Preconditioner}

In the context of solving linear systems, it can be observed that simple preconditioners such as Jacobi or matrix polynomial preconditioners are very good at reducing high frequency error, but struggle in reducing low frequency error components.
The reason is that only local information is taken into account by the preconditioner.
Here, locality is to be understood in the sense of distance in the matrix graph.

The idea of multigrid is to incorporate corrections computed on coarse approximations of the matrix \(A\), thereby enabling global information exchange.
Error components of low frequency are efficiently reduced on these coarser approximations by use of simple preconditioners (called \emph{smoothers} in this context).
Commonly, the operator \(A\) is coarsened often enough, so that a direct solve can be used on the coarsest level.
In many cases, particularly for systems \(A\) arising from discretization of partial differential equations, multigrid can achieve a constant condition number for the preconditioned linear system, independent of the problem size.
Since the application of this preconditioner is proportional in cost to the initial matrix, this implies that multigrid is an \emph{optimal} solver in such cases.

The setup of a multigrid preconditioner is more costly than the setup of Jacobi, and, based on experiments, also more expensive than a GMRES based polynomial preconditioner.
The main cost is the computation of coarse approximations of the matrix \(A\) which involves coarsening of the matrix graphs, resulting in a so called \emph{restriction} and \emph{prolongation} operators \(R\) and \(P\) and their triple products \(RAP\).
Since the entire hierarchy of approximations is computed using nothing but the original operator \(A\), this flavor of multigrid is called \emph{algebraic}.
(In contrast, \emph{geometric multigrid} uses rediscretization on coarser meshes.)
For symmetric matrices \(A\), \(R\) can be taken to be the transpose of \(P\), assuring that the coarse system \(RAP\) is also symmetric.

Sphynx relies on the \emph{smoothed aggregation algebraic multigrid} preconditioner implemented in the MueLu package~\cite{muelu} in Trilinos. (Plain aggregation is also an option.)


%% file: experiments.tex
\section{Experiments}
\label{sec:exp}

This section presents the experimental evaluation of Sphynx in several dimensions which cover
\begin{itemize}
\item both regular and irregular input graph types, 
\item using Jacobi, polynomial, and MueLu preconditioners in LOBPCG, 
\item using four different convergence tolerance values in LOBPCG,
\item using combinatorial, generalized, and normalized graph Laplacian eigenvalue problems,
\item the runtime of the bottleneck LOBPCG step among three major steps, and
\item comparison against the state-of-the-art partitioning tools ParMETIS~\cite{ParMetis}, XtraPuLP~\cite{xtrapulp}, and nvGRAPH's spectral partitioning tool~\cite{Naumov16}, which we call nvGRAPH in short.
\end{itemize}

We use two major performance metrics in our evaluations.  
These metrics are parallel partitioning running time (runtime) and the cutsize of the resulting partition.   
Here, cutsize refers to twice the number of cut edges, because each cut edge is counted twice by the two MPI processes that own its end vertices.
When comparing different variants of Sphynx, we sometimes use an additional metric: the number of iterations in LOBPCG.
We perform each experiment five times and report the geometric mean of the respective results. 

Another important quality metric is the imbalance on the part weights of the resulting partition.
In all compared methods, we use unit vertex weights.
We observe that all Sphynx, ParMETIS, and XtraPuLP variants achieve the smallest possible imbalance values, which is less than $1\%$ in most cases. 
Since those methods achieve the same imbalance in all instances, we omit the imbalance metric from evaluation of those methods.
However, nvGRAPH obtains larger imbalance values, therefore we include this metric in the nvGRAPH comparison.

We perform the experiments on Summit, the second fastest supercomputer on the TOP500 list as of June 2020.   
Each node of Summit contains six NVIDIA Volta V100 GPUs and two IBM Power9 processors with 42 physical CPU cores in total. 
Power9 processors are connected via dual NVLINK bricks that are capable of 25GB/s transfer rate in each direction. 
Each node contains 512 GB of DDR4 memory for use by the Power9 processors and 96 GB of High Bandwidth Memory for use by the GPUs. 
Nodes are interconnected in a nonblocking fat tree topology through a dual-rail EDR InfiniBand network.    

The primary purpose of this work is to study the performance of Sphynx on GPUs, so, we focus on the MPI+Cuda setting in our experiments.
In this setting, we use one GPU per MPI process.
Since there are six GPUs in each Summit node, we use six MPI processes per node.
We use Unified Virtual Memory (UVM) for device allocations, because Trilinos' GPU support relies on UVM.

For all partitioning instances with Sphynx, ParMETIS, and XtraPuLP, we use the same Trilinos-based driver which reads the input graph into a Tpetra CrsMatrix object with a 1D block distribution.
Using this driver, we call ParMETIS and XtraPuLP through their Zoltan2 interface.
We set the maximum allowable imbalance ratio of $1\%$ for all methods.

In our Trilinos build, we use CUDA 10 (version 10.1.243), GCC (version 7.4.0), Netlib's LAPACK (version 3.8.0), Spectrum MPI (version 10.3.1.2), ParMETIS (version 4.0.3),  METIS (version 5.1.0), XtraPuLP (version 0.3), and PuLP (version 0.2).
We set the environment variable CUDA\_LAUNCH\_BLOCKING as it is a general suggestion for Trilinos installations with Cuda.
Among the important Trilinos flags, we set 
\begin{itemize}
\item TPL\_ENABLE\_CUSPARSE=OFF and 
\item Tpetra\_ASSUME\_CUDA\_AWARE\_MPI=OFF.
\end{itemize}
We experimentally observed that setting those variables otherwise results in higher running times.

\subsection{Dataset}

Our test graphs consist of 8 regular and 12 irregular graphs, most of which are obtained from the SuiteSparse matrix collection~\cite{Davis2011}.
We excluded the small graphs with less than one million rows.
First and second halves of Table~\ref{t:ds} display our regular and irregular test graphs, respectively. 
For space efficiency, we use short names for dielFilterV2real, hollywood-2009, wikipedia-20070206, com-LiveJournal, and com-Friendster.
We symmetrized our test graphs by the $A+ A^T+I$ formulation. 
In case of a disconnected graph, we consider the largest component.

\begin{table}[]
\caption{Properties of test graphs.} 
\begin{center}
\scalebox{0.90}
{ 
\begin{tabular}{lrrrr} \toprule
& &  & \multicolumn{2}{c}{degree}  \\ \cmidrule(lr){4-5}
\multicolumn{1}{c}{name} & \multicolumn{1}{c}{\#vertices} & \multicolumn{1}{c}{\#edges} & \multicolumn{1}{c}{max} & \multicolumn{1}{c}{avg}   \\ \midrule
ecology1	&1,000,000	&4,996,000	&5	&5\\
dielFilter	&1,157,456	&48,538,952	&110	&42\\
thermal2	&1,227,087	&8,579,355	&11	&7\\
Bump\_2911	&2,852,430	&127,670,910	&195	&45\\
Queen\_4147	&4,147,110	&329,499,284	&81	&79\\
$100^3$	&1,000,000	&26,463,592	&27	&26\\
$200^3$	&8,000,000	&213,847,192	&27	&27\\
$400^3$	&64,000,000	&1,719,374,392	&27	&27\\ \midrule
hollywood	&1,069,126	&113,682,432	&11,468	&106\\
com-Orkut	&3,072,441	&237,442,607	&33,314	&77\\
wikipedia	&3,512,462	&88,261,228	&187,672	&25\\
cit-Patents	&3,764,117	&36,787,597	&794	&10\\
LiveJournal	&3,997,962	&73,360,340	&14,816	&18\\
wb-edu	&8,863,287	&97,233,789	&25,782	&11\\
uk-2005	&39,252,879	&1,602,132,663	&1,776,859	&41\\
it-2004	&41,290,577	&2,096,240,367	&1,326,745	&51\\
twitter7	&41,652,230	&2,446,678,322	&2,997,488	&59\\
Friendster	&65,608,366	&3,677,742,636	&5,215	&56\\
FullChip	&2,986,914	&26,621,906	&2,312,481	&9\\
circuit5M	&5,555,791	&59,519,031	&1,290,501	&11\\
\bottomrule
\end{tabular}
}
\end{center}
\label{t:ds}
\end{table}


All of our regular graphs are obtained from a mesh.
The first 5 regular graphs in the table are \textit{real} and obtained from the SuiteSparse collection by querying "2D/3D" problem type.   
The last 3 regular graphs are \textit{synthetic} and created by Trilinos package Galeri by setting "Brick3D" problem type which corresponds to a 27-point stencil discretization of the standard Laplacian differential operator on a uniform mesh.
In the naming of these 3D synthetic meshes, $x^3$ refers to deploying $x$ elements along each dimension, for $x\in\{100,200,400\}$.

All of our irregular graphs are obtained from the SuiteSparse collection. 
The first 10 of them correspond to a \textit{web} graph or a \textit{social network}, whereas the last two correspond to a \textit{circuit simulation} graph.
Each irregular graph has at least one high-degree vertex.

As will be clear throughout this section, the performance optimizations needed for regular and irregular graphs are different.
Hence, Sphynx's default settings are different for these two types.
Our users might be oblivious to the structure of their input graph, so, Sphynx has a simple mechanism to determine the graph type as follows.   
The input graph is identified as regular only if the ratio of maximum to average degree is smaller than or equal to 10, and it is identified as irregular, otherwise. 
In our regular test graphs, minimum and maximum values for this ratio are 1.00 and 4.36, respectively.  
In our irregular test graphs, the minimum and maximum values are 81 and 259,454, respectively.

\subsection{Framework}

In Sections \ref{sec:init} and \ref{sec:mueluparams}, we discuss the default settings for vector initialization and MueLu parameters, respectively.
In Section~\ref{sec:default}, we present a decision flow diagram to show the default values of the parameters that are more critical for performance. 
The results and discussions given in Section~\ref{sec:results} also include the sensitivity analysis for those parameters. 

\subsubsection{Initial Vectors for LOBPCG}
\label{sec:init}
Depending on the graph type, we use different vector initialization methods: random or piecewise constant initial vectors.
While random initial vectors span a subspace of size \(d\) in practice, the procedure to choose piecewise constant vectors is as follows.
The first vector is chosen to be constant equal to one, taking into account that we already know this to be an eigenvector with eigenvalue zero.
We then divide the global index space into \(d\) blocks, and set the other \(d-1\) vectors to be equal to one on an assigned block and zero otherwise.
Since we only select \(d-1\) out of the \(d\) blocks, the vector of all ones is not in the span of these vectors.
We observed in our preliminary experiments that random and piecewise constant vectors result in smaller cutsize on regular and irregular graphs, respectively.
Hence, we use random vectors for regular graphs and piecewise constant vectors for irregular graphs.

\subsubsection{MueLu Preconditioner Parameters}
\label{sec:mueluparams}
MueLu allows its users to set a wide range of parameters.
In this section, we cover only some of them and refer the reader to the manual for other parameters~\cite{muelu}.
By default, MueLu uses one sweep of symmetric Gauss-Seidel for smoothing and SuperLU for coarse system solver.
Chebyshev smoothing is better suited to the use on GPUs, since it only relies on the SpMV kernel.
During the setup of the preconditioner, eigenvalue estimates are obtained using 10 steps of power iteration and an eigenvalue ratio of 7.
Based on the results of experiments, we use a degree 3 Chebyshev polynomial.

The restriction matrix $R$ is equal to the prolongation $P^T$ in the $RAP$ operation, because the Laplacian matrix (which corresponds to $A$ in $RAP$) is symmetric.
This feature enables us to use the implicit restriction construction that avoids the communication performed during the explicit transpose.

For irregular graphs, we further improved the set of MueLu parameters we use. 
First, we use unsmoothed aggregation to prevent the coarse matrices from becoming too dense.
Second, we set a drop tolerance of 0.4 that removes small values from the graph considered in the coarsening with the same purpose of preventing the coarse matrices from becoming dense.
Third, we limit the number of levels to 5.
This is because on irregular graphs, coarse systems sometimes do not get sufficiently small, so coarsening may take many levels with  large graphs, which in turn causes a memory problem.
Lastly, we use Chebyshev for the coarse system solver to avoid the direct solver cost in the case of a coarse system which is not sufficiently small.
Here, we use the same Chebysev settings as in the smoother settings except that we perform 100 steps of power iteration.

\input{exp-defaults}

\subsection{Results}
\label{sec:results}

\input{exp-tolerance}
\input{exp-problem}
\input{exp-lobpcgtime}
\input{exp-preconditioner}
\input{exp-stateoftheart}


%% file: exp-defaults.tex
\subsubsection{Sphynx Default Settings}
\label{sec:default}

\begin{figure*}[t]
\centering
\includegraphics[width=6in]{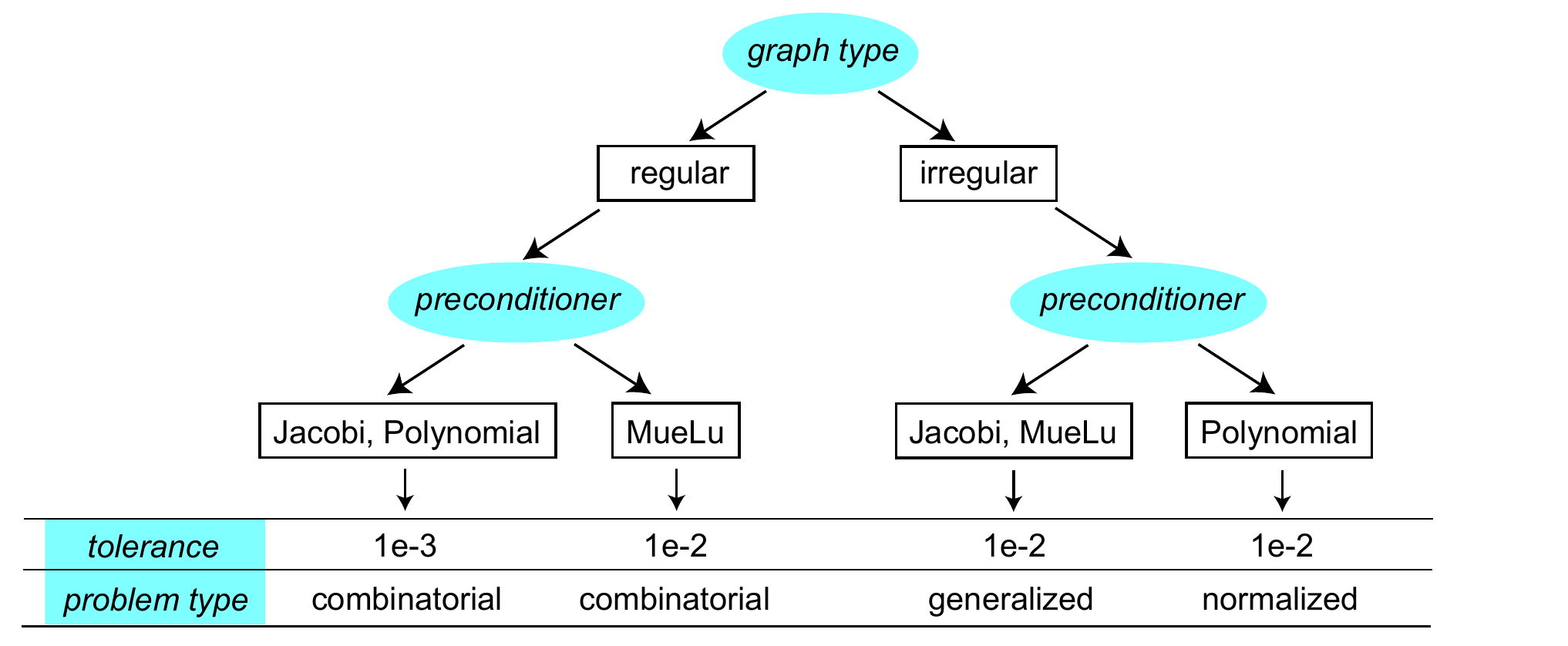}
\caption{The decision flow diagram that shows the default values of LOBPCG convergence tolerance and eigenvalue problem type.}
\label{fig:diagram}
\end{figure*}

This section describes the decision flow for setting the default values of the following parameters in Sphynx:
\begin{itemize}
\item LOBPCG convergence tolerance and
\item eigenvalue problem type.
\end{itemize}
Since our focus is to study the performance of Sphynx with different preconditioners, we keep all preconditioners in the results presented in the following sections, hence, we are not treating the preconditioner type as a parameter for now.

Figure~\ref{fig:diagram} shows a diagram of the decision flow for the default values of tolerance and problem type.
The decisions shown in this diagram are based on the experiments which will be described in detail in Sections~\ref{sec:tol}~and~\ref{sec:pr}.
We introduce this diagram before the experimental results, because the following sections will refer to the default settings for the parameters that are not mentioned in those sections.

As seen in Figure~\ref{fig:diagram}, Sphynx takes different actions for different graph types and preconditioners.
For regular graphs, it uses the combinatorial problem by default, regardless of the preconditioner type. 
However for irregular graphs, it uses the generalized problem for the Jacobi and MueLu preconditioners and the normalized problem for the polynomial preconditioner.
On regular graphs, with the Jacobi and polynomial preconditioners, Sphynx uses 1e-3 as the default value for tolerance.
With the MueLu preconditioner on regular graphs, the default value is 1e-2.
On irregular graphs, the default value is 1e-2. regardless of the preconditioner type.

%% file: exp-tolerance.tex
\subsubsection{LOBPCG Convergence Tolerance}
\label{sec:tol}
\begin{figure}
    \begin{center}
     	\subfloat{\includegraphics[scale=0.60]{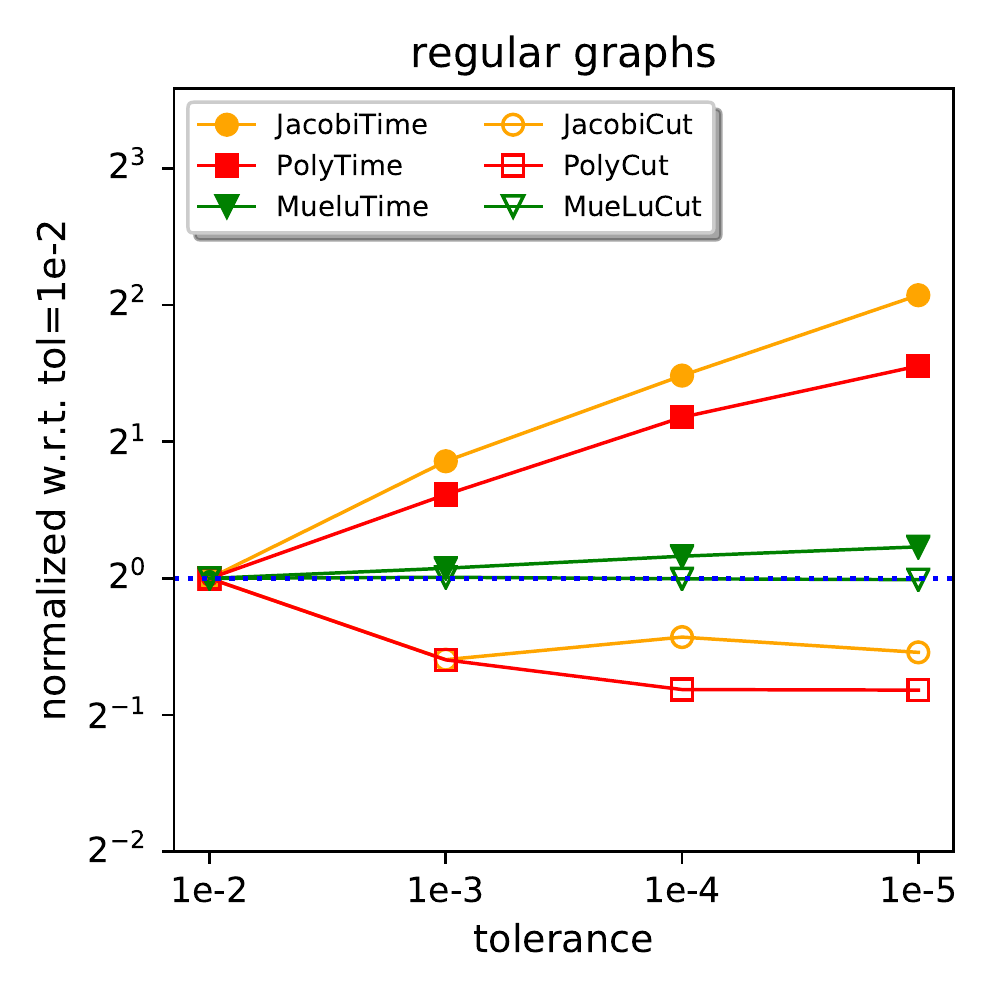}}  \\
      	\subfloat{\includegraphics[scale=0.60]{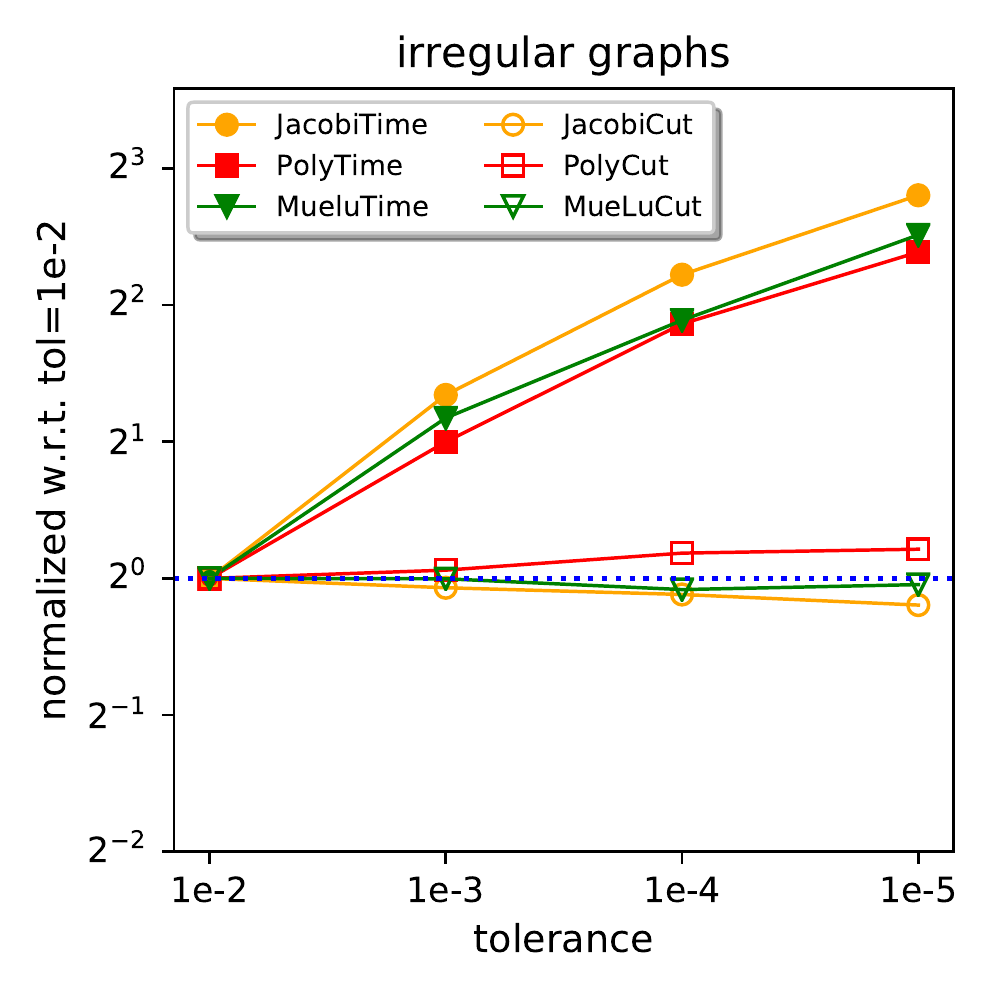}}
    \end{center}
\caption{Results (runtime and cutsize) with different tolerance values normalized with respect to the results at tolerance 1e-2, averaged over regular/irregular graphs with geometric mean.}
\label{fig:tol}
\end{figure}

In this section, we evaluate the effect of the LOBPCG convergence tolerance on the results.
In our evaluation, we consider four different values: 1e-2, 1e-3, 1e-4, and 1e-5.
We conducted these experiments on 24 GPUs and we set the number of parts $K$ to 24.
The top and bottom plots in Figure~\ref{fig:tol} display the average results on regular and irregular graphs, respectively. 
In these plots, each color and marker shape represents using a different preconditioner in Sphynx.
Orange circle, red square and green triangle represent Jacobi, polynomial, and MueLu preconditioners, respectively. 
Filled and empty markers represent runtime and cutsize, respectively.

For a given preconditioner $P$ and tolerance $t$, the y value on the runtime curve denotes the Sphynx result with tolerance $t$ normalized with respect to the Sphynx result with tolerance \mbox{1e-2}.
The same is also true for the cutsize curves.
The values seen in the plots are the geometric means of the normalized results over all regular/irregular graphs.
Note that all curves coincide at $y=1$ for tolerance 1e-2, simply because normalizations are performed with the results at 1e-2.
\textit{It is important to highlight that each line in these plots should be considered individually.}
That is, the line corresponding to one preconditioner being below the line corresponding to another preconditioner does not imply anything about how those preconditioners perform with respect to each other.     
    
As seen in the top plot of Figure~\ref{fig:tol}, using different tolerance values only slightly affects the performance of Sphynx with the Muelu preconditioner on regular graphs.
Average cutsize does not change at all, while runtime slightly increases with decreasing tolerance values.
On the other hand, the performance of Sphynx with the Jacobi and polynomial preconditioners is drastically affected by using different tolerance values.
At tolerance 1e-5, the runtimes with Jacobi, polynomial and MueLu are 4.2x, 2.9x, and 1.2x larger than those at tolerance 1e-2, respectively.
The cutsizes with Jacobi, polynomial and MueLu at 1e-5 are 31\%, 43\%, and 1\% smaller than those at 1e-2, respectively.
Therefore, for regular graphs, we set the default value for tolerance as follows.
For polynomial and Jacobi, we set it to 1e-3, because the cutsize at 1e-3 is 34\% smaller than the cutsize at 1e-2.
Note that for these two preconditioners, decreasing tolerance further does not pay off when compared to the increase in the runtime after 1e-3. 
For MueLu, we set the default value for tolerance to 1e-2, because the cutsize performance is very similar for all values considered, while 1e-2 is the fastest option.
 
On some irregular graphs, LOBPCG with Jacobi and polynomial preconditioners failed due to a Cholesky factorization error (breakdown due to singular matrix). This rare breakdown is 
caused by linearly dependent vectors in LOBPCG and is a 
numerical issue in Anasazi so beyond the scope of Sphynx.
Note none of the runs with MueLu failed in our experiments, so using preconditioner MueLu proved to be the most robust alternative among the three options.
To have a fair comparison of using different tolerance values in the bottom plot of Figure~\ref{fig:tol}, for the graphs where at least one run failed using a preconditioner, we removed all instances of those graphs from the results of that preconditioner.
Specifically, we removed it-2004 and twitter7 from Jacobi's results and uk-2005, it-2004, and circuit5M from polynomial's results.
This error only occurs on irregular graphs with a tolerance smaller than 1e-2, so the default Sphynx version does not have this issue.
 
As seen in the bottom plot of Figure~\ref{fig:tol}, using different tolerance values only slightly changes the cutsize on irregular graphs with each preconditioner.
On the other hand, the runtime drastically increases with decreasing tolerance for all preconditioners.
At tolerance 1e-5, the runtimes with Jacobi, polynomial and MueLu are 7.0x, 5.2x, and 5.7x larger than those at tolerance 1e-2, respectively.
The cutsizes with Jacobi and MueLu at 1e-5 are 13\% and 3\% smaller than those at 1e-2, respectively, whereas the cutsize is 16\% larger with polynomial.
Note that this cutsize increase aligns with the cutsize increase mentioned in~\cite{AcerBomanRajamanickam-Sphynx}, which uses the normalized problem.
So, we speculate that the decreasing tolerance causes an increase in the cutsize only for the normalized problem.    
In conclusion, for irregular graphs, we set the default value of tolerance to 1e-2, because it is the fastest option with a reasonable cutsize.

%% file: exp-problem.tex
\subsubsection{Eigenvalue Problem}
\label{sec:pr}
In this evaluation, we consider three eigenvalue problems: combinatorial, generalized, and normalized graph Laplacian problems.
As in the previous subsection, we conducted these experiments on 24 GPUs and we set the number of parts $K$ to 24.

%
%
%

\begin{table}
  \caption{Average results of generalized and normalized eigenvalue problems normalized w.r.t. those of combinatorial eigenvalue problem.} 
  \begin{center}
 \scalebox{0.87}
 {    
    \begin{tabular}{l l r r r r r r}
    \toprule
 & & \multicolumn{3}{c}{generalized} & \multicolumn{3}{c}{normalized} \\ \cmidrule(lr){3-5}  \cmidrule(lr){6-8}
 &precond. &\#iters & time & cut & \#iters & time & cut \\ \midrule 
\multirow{3}{*}{\rotatebox{90}{regular}}	&Jacobi	&0.68	&0.81	&1.15	&0.38	&0.43	&2.26\\
	&Polynomial	&0.59	&0.73	&1.21	&0.36	&0.54	&2.45\\
	&MueLu	&0.60	&0.99	&1.12	&0.44	&0.95	&2.20\\ \midrule
\multirow{3}{*}{\rotatebox{90}{irregular}}	&Jacobi	&0.71	&0.75	&0.83	&0.19	&0.26	&1.36\\
	&Polynomial	&0.36	&0.36	&0.84	&0.02	&0.02	&0.83\\
	&MueLu	&0.66	&0.71	&0.90	&0.15	&0.31	&1.68\\	
    \bottomrule
    \end{tabular}
 }
  \end{center}
  \label{t:pr}
\end{table}

Table~\ref{t:pr} displays the average results of the generalized and normalized eigenvalue problems normalized with respect to those of the combinatorial eigenvalue problem.
As in the previous section, for each graph with at least one failing run with a preconditioner, we exclude that graph from the corresponding preconditioner's results. 
Specifically, we removed wb-edu, uk-2005, it-2004, twitter7, FullChip, and circuit5M from polynomial's results.

On regular graphs, in the generalized problem, the number of LOBPCG iterations is 32\%, 41\%, and 40\% smaller compared to the combinatorial problem, which leads to a reduction of 19\%, 27\%, and 1\%  in runtime for Jacobi, polynomial, and MueLu, respectively.
Similarly in the normalized problem, the number iterations is 62\%, 64\%, and 56\% smaller, which leads to a reduction of 57\%, 46\%, and 5\% in runtime for Jacobi, polynomial, and MueLu, respectively.
For MueLu, the 40\%-56\% reduction in the number of iterations translates into a slight improvement of 1\%-5\% in runtime, unlike other preconditioners.
This is because MueLu's setup time is a large portion of the overall Muelu-enabled Sphynx time on regular graphs, while other preconditioners do not incur a significant setup cost.
While the runtimes of the generalized and normalized problems are smaller compared to the combinatorial problem, their cutsizes are larger.
The cutsize of the generalized problem is 15\%, 21\%, and 12\% larger, whereas the cutsize of the normalized problem is 126\%, 145\%, and 120\% larger, for Jacobi, polynomial, and MueLu, respectively.
Therefore, since the combinatorial problem consistently results in the smallest cutsize for all preconditioners, we set the default eigenvalue problem in Sphynx to the combinatorial problem for regular graphs.

On irregular graphs, in the generalized problem, the number of LOBPCG iterations is 29\%, 64\%, and 34\% smaller compared to the combinatorial problem, which leads to a reduction of 25\%, 64\%, and 29\%  in runtime for Jacobi, polynomial, and MueLu, respectively.
Similarly in the normalized problem, the number iterations is 81\%, 98\%, and 85\% smaller, which leads to a reduction of 74\%, 98\%, and 69\% in runtime for Jacobi, polynomial, and MueLu, respectively.
MueLu results in more iterations on irregular graphs than regular graphs, so the translation of the reduction in the number of iterations into a reduction in runtime is more apparent.
Unlike the case with regular graphs, the cutsize of the generalized problem is smaller than the cutsize of the combinatorial problem, by 17\%, 16\%, and 10\%, for Jacobi, polynomial, and MueLu, respectively.
However with the normalized problem, the cutsize is larger than the cutsize of the combinatorial problem with Jacobi and MueLu by 36\% and 68\%, respectively, while it is smaller by 17\% with polynomial.
Therefore, for Jacobi and MueLu, we set the default to the generalized problem, because it is the fastest option with the smallest cutsize. 
However for polynomial, we set the default to the normalized problem, because even though both generalized and normalized problems obtain the smallest cutsizes, the normalized problem is 18x faster than the generalized problem.

%% file: exp-lobpcgtime.tex
\subsubsection{LOBPCG Runtime}
In this section, we analyze the LOBPCG runtime compared to the overall Sphynx runtime. 
Recall that Sphynx has the following major steps: 
\begin{enumerate}[(i)]
\item creating the Laplacian matrix $L$ (and degree matrix $D$ for the generalized problem),
\item calling the LOBPCG eigensolver, and
\item calling the multi-jagged (MJ) algorithm.
\end{enumerate}
In most cases, the runtimes of steps (i) and (iii) are very small compared to the runtime of step (ii), so, we only focus on the LOBPCG runtime.
Here, we use the results of the experiments in the previous subsection.

On average, more than 87\% of the Sphynx runtime is spent in the LOBPCG eigensolver. 
On regular graphs, on average, LOBPCG takes 97\%, 92\%, and 91\% of the overall Sphynx execution for Jacobi, polynomial, and MueLu, respectively. 
On irregular graphs, these values become 97\%, 87\%, and 98\%, for Jacobi, polynomial, and MueLu, respectively.

Compared to the previous work~\cite{AcerBomanRajamanickam-Sphynx}, the runtime percentage for step (ii) has increased. This is mostly because Sphynx now defaults to the combinatorial or generalized problem in most cases.
Recall from Table~\ref{t:pr} that the combinatorial and generalized problems have more iterations than the normalized problem (approximately 2x-3x more iterations on regular graphs and 4x-50x more iterations on irregular graphs). Hence, the percentage of the LOBPCG runtime increased due to the increase in the number of iterations. 
Note that the smallest LOBPCG percentage is observed with polynomial preconditioner on irregular graphs because Sphynx uses the normalized problem in that case.

%% file: exp-preconditioner.tex
\subsubsection{Preconditoner}
This section evaluates the performance of Sphynx with Jacobi, polynomial, and MueLu preconditioners.
In Sphynx, we treat preconditioners as an available option which the user can select,
depending on whether or not they are enabled in the respective Trilinos installation.
However, this evaluation covers the case when all preconditioners are enabled and the user does not have a preference on which preconditioner to use.
 
\begin{table}[h]
  \caption{Sphynx's performance with Jacobi, polynomial, and MueLu preconditioners on 24 GPUs. Jacobi results are actual values, whereas the runtime and cutsize results of the other preconditioners are given as speedup and cutsize improvement (reduction) factors over Jacobi, respectively.} 
  \vspace{-2ex}
  \begin{center}
 \scalebox{0.83}
 {    
    \begin{tabular}{l r r r r r r}
    \toprule
& & \multicolumn{2}{c}{speedup} & & \multicolumn{2}{c}{cutsize imp.} \\
 &  \multicolumn{1}{c}{time (s)} & \multicolumn{2}{c}{w.r.t. Jacobi} & \multicolumn{1}{c}{cutsize} & \multicolumn{2}{c}{w.r.t. Jacobi} \\ \cmidrule(lr){2-2} \cmidrule(lr){3-4} \cmidrule(lr){5-5} \cmidrule(lr){6-7}    
graph	&Jacobi	&Poly.	&MueLu	&Jacobi	&Poly.	&MueLu\\ \midrule
ecology1	&3.23	&3.23x	&3.40x	&44.2K	&0.77x	&1.71x\\
dielFilter	&4.57	&1.38x	&2.05x	&2.2M	&1.00x	&0.98x\\
thermal2	&4.11	&1.82x	&1.95x	&58.0K	&1.07x	&1.31x\\
Bump\_2911	&3.97	&1.76x	&1.99x	&5.8M	&1.00x	&1.00x\\
Queen\_4147	&4.47	&2.39x	&1.64x	&14.4M	&1.00x	&1.00x\\
100$^3$	&2.11	&2.45x	&1.61x	&1.3M	&1.03x	&0.99x\\
200$^3$	&5.39	&2.40x	&2.42x	&5.0M	&0.97x	&0.98x\\
400$^3$	&37.00	&2.43x	&5.72x	&19.5M	&0.96x	&0.97x\\ 
\textbf{geomean}	&	&\textbf{2.17x}	&\textbf{2.37x}	&	&\textbf{0.97x}	&\textbf{1.09x}\\ \midrule
hollywood	&3.07	&1.13x	&0.64x	&61.0M	&1.01x	&1.02x\\
com-Orkut	&4.73	&0.57x	&0.53x	&137.1M	&1.01x	&1.02x\\
wikipedia	&7.65	&1.03x	&0.42x	&66.0M	&1.06x	&1.01x\\
cit-Patents	&4.69	&0.85x	&0.50x	&14.8M	&0.80x	&1.04x\\
LiveJournal	&5.13	&1.44x	&0.53x	&41.2M	&0.94x	&1.07x\\
wb-edu	&2.99	&2.58x	&0.57x	&5.2M	&0.15x	&1.07x\\
uk-2005	&37.02	&3.59x	&0.42x	&150.4M	&0.19x	&1.85x\\
it-2004	&36.47	&4.40x	&0.50x	&57.9M	&0.06x	&1.05x\\
twitter7	&219.17	&3.14x	&0.50x	&1.9B	&1.16x	&0.97x\\
Friendster	&166.42	&1.82x	&1.11x	&2.5B	&1.21x	&0.97x\\
FullChip	&22.43	&0.72x	&0.44x	&19.1M	&0.92x	&0.93x\\
circuit5M	&14.75	&2.39x	&0.38x	&35.0M	&0.93x	&1.01x\\
\textbf{geomean}	&	&\textbf{1.62x}	&\textbf{0.52x}	&	&\textbf{0.58x}	&\textbf{1.07x}\\
    \bottomrule 
    \end{tabular}
 }
  \end{center}
  \label{t:prec}
  \vspace{-1ex}
\end{table}

\begin{table}[h]
  \caption{Average number of LOBPCG iterations in Sphynx on 24 GPUs. } 
  \vspace{-2ex}
  \begin{center}
 \scalebox{0.83}
 {    
 \begin{tabular}{l r r r}
    \toprule
	&Jacobi	&Poly.	&MueLu\\ \midrule
regular	&358	&22	&4\\
irregular	&110	&2	&29 \\
    \bottomrule 
    \end{tabular}
 }
  \end{center}
  \label{t:niter}
  \vspace{-2ex}
\end{table}

Table~\ref{t:prec} displays the runtime and cutsize performance of Sphynx with each preconditioner on each test graph on 24 GPUs, while Table~\ref{t:niter} displays the resulting number of iterations in these experiments averaged over regular and irregular graphs. 
In Table~\ref{t:prec}, Jacobi columns display actual results, whereas the other columns display speedup values (columns 3-4) and cutsize improvement (reduction) factors (columns 6-7) with respect to Jacobi, where values greater than 1 signify better results.
On regular graphs, polynomial and MueLu obtain average speedups of 2.17x and 2.37x compared to Jacobi, respectively. 
This is mostly because the average number of iterations with Jacobi, polynomial, and MueLu on regular graphs are 358, 22, and 4, respectively. 
In terms of cutsize, polynomial causes a 3\% degradation (increase), whereas MueLu obtains 9\% improvement (reduction) on average.
On irregular graphs, polynomial obtains an average speedup of 1.62x, whereas MueLu causes an average slowdown with 0.52x. 
This is mostly because the average iteration counts of Jacobi, polynomial, and MueLu on irregular graphs are 110, 2, and 29, respectively. 
In terms of cutsize, polynomial causes an average degradation with 0.58x (mostly due to wb-edu, uk-2005, and it-2004), whereas MueLu obtains an average improvement of 1.07x.
Consequently, when all three preconditioners are available, we favor using MueLu on regular graphs and polynomial on irregular graphs. 

When we consider the computation and data movement costs with the three preconditioners, per iteration costs are the cheapest with Jacobi due to the simple diagonal scaling and the most expensive with MueLu due to its multilevel nature.
Nonetheless, our experimental results show that the number of iterations is a more important factor in determining the runtime performance compared to the per iteration cost. This is what would be expected when the total time is more important than the per iteration cost.
For example, the runtimes with MueLu and polynomial being the smallest on regular and irregular graphs, respectively, is explained by their drastically smaller number of iterations. 

%% file: exp-stateoftheart.tex
\subsubsection{Comparison against State-of-the-art Partitioners}
This section compares the performance of Sphynx against three state-of-the-art partitioners: ParMETIS~\cite{ParMetis}, XtraPuLP~\cite{xtrapulp}, and nvGRAPH's spectral partitioning~\cite{Naumov16}, which we call nvGRAPH in short.
ParMETIS~\cite{ParMetis} and XtraPuLP~\cite{xtrapulp} are both MPI based but do not run on GPUs.
We include them in the comparison because ParMETIS is the most commonly used partitioner in the community, while XtraPuLP, a more recent partitioner, has proven successful on irregular graphs.  
nvGRAPH\cite{Naumov16}, on the other hand, runs on GPUs, however, does not support any distributed-memory parallelism.  
We include it in the comparison to give users an idea about Sphynx's single GPU performance with respect to nvGRAPH.

\paragraph{Application-friendly comparison}
We first compare Sphynx against ParMETIS and XtraPuLP from the perspective of a distributed application.
Consider a Trilinos-based, GPU-enabled, distributed application, where a sparse matrix has already been read into MPI processes with 1D block distribution.
The application calls the partitioner on the fly and redistributes the matrix according to the resulting partition.
This means the initial read operation, the partitioning, and the application itself all belong to the same execution, so the number of MPI processes will remain the same during each step.  
To exemplify this use case, we ran all compared partitioners on 24 MPI processes on 4 nodes.
In this setting, each node gets 6 MPI processes so that the application uses one GPU per MPI process, which is a Trilinos requirement.
ParMETIS is an MPI-only code, so the 42-6=36 CPU cores of each node will remain idle during the partitioning.
XtraPuLP is an MPI+OpenMP code, so we use 7 OpenMP threads per MPI process in order to utilize all CPU cores on the nodes.
Similar to ParMETIS, Sphynx uses only 6 of the 42 CPU cores, but it fully utilizes all 6 GPUs on each node. 

We used the same Trilinos-based driver for all three partitioners and called ParMETIS and XtraPuLP through their Zoltan2 interfaces.
We set the maximum allowed imbalance ratio to 1\%.
As suggested by the previous subsection, Sphynx uses MueLu on regular graphs and polynomial on irregular graphs.
Table~\ref{t:state} displays the runtime and cutsize results.
The columns under Sphynx display the actual values, whereas the other results are normalized with respect to Sphynx's results.  

\begin{table}
  \caption{Comparison of Sphynx against ParMETIS and XtraPuLP on 24 MPI processes.} 
  \vspace{-2ex}
  \begin{center}
 \scalebox{0.88}
 {    
    \begin{tabular}{l r r r r r r}
    \toprule
	&\multicolumn{2}{c}{actual values}	& \multicolumn{4}{c}{normalized w.r.t. Sphynx}\\ \cmidrule(lr){2-3} \cmidrule(lr){4-7}
	&\multicolumn{2}{c}{Sphynx}	&\multicolumn{2}{c}{ParMETIS}	&\multicolumn{2}{c}{XtraPuLP}	\\ \cmidrule(lr){2-3} \cmidrule(lr){4-5} \cmidrule(lr){6-7}
	&time	&cut	&time	&cut	&time	&cut\\ \midrule
ecology1	&0.95	&26.0K	&0.07	&0.84	&0.19	&22.00\\
dielFilter	&2.23	&2.2M	&0.41	&0.58	&0.23	&2.25\\
thermal2	&2.11	&44.4.K	&0.13	&0.84	&0.11	&17.95\\
Bump\_2911	&2.00	&5.8M	&0.38	&0.84	&0.32	&3.37\\
Queen\_4147	&2.73	&14.4M	&0.67	&0.85	&0.56	&3.83\\
100$^3$	&1.31	&1.3M	&0.17	&0.80	&0.12	&3.14\\
200$^3$	&2.23	&5.1M	&0.60	&0.88	&0.53	&6.18\\
400$^3$	&6.47	&20.3M	&1.61	&0.91	&1.56	&12.04\\
\textbf{geomean}&	&	&\textbf{0.33}	&\textbf{0.81}	&\textbf{0.31}	&\textbf{6.36}\\ \midrule
hollywood	&2.72	&60.4M	&13.39	&0.55	&1.30	&0.75\\
com-Orkut	&8.31	&135.3M	&13.68	&0.61	&0.88	&0.83\\
wikipedia	&7.43	&62.4M	&23.41	&0.55	&1.29	&0.81\\
cit-Patents	&5.53	&18.4M	&1.57	&0.25	&0.44	&0.56\\
LiveJournal	&3.56	&43.9M	&8.34	&0.38	&1.20	&0.54\\
wb-edu	&1.16	&35.4M	&8.92	&0.02	&1.07	&0.07\\
uk-2005	&10.30	&802.3M	&	DNF & DNF	&3.39	&0.19\\
it-2004	&8.29	&989.7M	&	DNF & DNF	&3.64	&0.07\\
twitter7	&69.71	&1.6B	&	DNF & DNF	&2.84	&1.20\\
Friendster	&91.67	&2.1B	&	DNF & DNF	&2.99	&0.86\\
FullChip	&30.97	&20.8M	&223.51	&0.41	&0.11	&0.54\\
circuit5M	&6.18	&37.4M	&966.67	&0.40	&1.31	&0.84\\
\textbf{geomean}	&	&	&\textbf{23.95}	&\textbf{0.30}	&\textbf{1.24}	&\textbf{0.45}\\
	
    \bottomrule
    \end{tabular}
 }
  \end{center}
  \vspace*{-0.5em}
  \footnotesize{DNF: Did Not Finish in 2 hours}
  \label{t:state}
\end{table}

As seen in Table~\ref{t:state}, on regular graphs, both ParMETIS and XtraPuLP are faster than Sphynx, by 67\% and 69\% on average, respectively.
However, Sphynx obtains smaller cutsize than XtraPuLP by 6.36x, whereas ParMETIS obtains smaller cutsize than Sphynx by 19\% on average.
These results confirm ParMETIS' superiority on regular graphs, but Sphynx and XtraPuLP get very close to ParMETIS in terms of cutsize and runtime, respectively.
On irregular graphs, Sphynx is faster than ParMETIS by 23.95x on average, however, ParMETIS obtains 30\% of the cutsize obtained by Sphynx on average.
Note that the ParMETIS execution did not finish on the four largest graphs (uk-2005, it-2004, twitter7, and Friendster) within the 2-hour limit, while both Sphynx and XtraPuLP executions finished on all test problems.
So the average performance numbers of ParMETIS exclude those instances.
Sphynx on irregular graphs is also faster than XtraPuLP, by 1.24x, however, XtraPuLP obtains 45\% of the cutsize obtained by Sphynx, on average.
These results suggest that Sphynx is a good GPU-based alternative on irregular graphs as it provides a faster execution at the expense of increased cutsize.

\paragraph{Comparison with better resource utilization}
Next, we compare Sphynx against ParMETIS and XtraPuLP on configurations that utilize the resources on 4 nodes better than the application-friendly comparison.
In particular, we ran ParMETIS on 4x42=168 MPI processes so that each CPU core on the 4 nodes gets an MPI process.
We ran XtraPuLP with two extreme settings on 4 nodes: with 168 MPI processes (each MPI process gets one thread), and with 4 MPI processes (each MPI process gets 42 threads) according to~\cite{private}.
Note that all of the CPU cores on the 4 nodes are utilized in these new settings of the baseline algorithms. 
Also note that this is just for a partitioner comparison and applications may not be able to take advantage of these additional cores in their typical runs.
For Sphynx, we used the same setting with 24 GPUs on the 4 nodes, where each MPI process gets a GPU.
We keep the partitioning problem the same, so the desired number of parts is still 24 for all settings.

\begin{table}
  \caption{ Average normalized results of ParMETIS (on 168 MPI processes) and XtraPuLP (on 168 and 4 MPI processes) with respect to Sphynx results (on 24 MPI processes). All cores on 4 nodes are utilized in the baseline algorithms. XtraPuLP with 4 MPI processes uses 42 OpenMP threads per MPI process.} 
  \vspace{-2ex}
  \begin{center}
 \scalebox{0.88}
 {    
    \begin{tabular}{l r r r r r r}
    \toprule
	& \multicolumn{2}{c}{ParMETIS168}	& \multicolumn{2}{c}{XtraPuLP168} & \multicolumn{2}{c}{XtraPuLP4}\\ \cmidrule(lr){2-3} \cmidrule(lr){4-5} \cmidrule(lr){6-7}
	& time	&cut	&time	&cut	&time	&cut\\ \midrule
	regular & 0.10 & 0.80 & 0.32 & 6.47 & 1.38 & 6.27 \\
	irregular & 18.20 & 0.29 & 0.98& 0.60& 2.97 & 0.41 \\
	    \bottomrule
    \end{tabular}
 }
  \end{center}
  \label{t:util}
  \vspace{-1ex}
\end{table}

Table~\ref{t:util} shows the average normalized results of the baseline algorithms with the above-mentioned settings with respect to Sphynx's results.
When ParMETIS uses 168 processes instead of 24, its normalized runtime reduces from 0.33 to 0.10 on regular graphs and from 23.95 to 18.20 on irregular graphs (see Table~\ref{t:state} for the average values on 24 processes). 
So with this new setting, ParMETIS is 10x faster on regular graphs and 18.10x slower on irregular graphs compared to Sphynx.
Note that there is no significant change in the cutsize obtained by ParMETIS on different settings.  
When XtraPuLP uses 168 process (with one thread per process) instead of 24 processes (with 7 thread per process), its normalized runtime does not change significantly on regular graphs but decreases from 1.26 to 0.98 on irregular graphs.
Note that with this setting, its normalized cutsize increases from 6.32 to 6.47 on regular graphs and from 0.44 to 0.60 on irregular graphs.
In the last setting, where XtraPuLP uses 4 processes (with 42 threads per process) instead of 24 processes, its runtime increases from 0.31 to 1.38 on regular graphs and from 1.26 to 2.97 on irregular graphs. 
So in this setting, Sphynx is faster than XtraPuLP on both regular and irregular graphs, by 38\% and 197\%, respectively.
The cutsize with the new setting slightly decreases from 6.32 to 6.27 on regular graphs and from 0.44 to 0.41 on irregular graphs.
These results suggest that ParMETIS and XtraPuLP on 168 MPI processes perform better than Sphynx on 24 MPI processes on regular and irregular graphs, respectively.

\paragraph{Comparison against nvGRAPH on a single GPU}

Since nvGRAPH runs on a single GPU, we ran Sphynx on a single GPU in these experiments as well. 
We set the number of parts to 24 (for both methods). 
Unlike Sphynx, nvGRAPH does \textit{not} use a default value for the number of eigenvalues to be used and relies on the user to provide it. 
We set this value to 4 in nvGRAPH experiments because Sphynx also uses 4 eigenvectors when partitioning into 24 parts due to $\lfloor \log_2(24)\rfloor = 4$. 
We set the algorithm in nvGRAPH to NVGRAPH\_BALANCED\_CUT\_LOBPCG and used default settings for the rest of the parameters. 

\begin{table}
  \caption{nvGRAPH results when partitioning into 24 parts and the improvements obtained by Sphynx over nvGRAPH on a single GPU. Sphynx improvement values larger than one signify that Sphynx obtains smaller (better) results. ``imb'' corresponds to imbalance=maximum/average part weight in the resulting partition.} 
  \vspace{-2ex}
  \begin{center}
 \scalebox{0.88}
 {    
    \begin{tabular}{l r r r r r r}
    \toprule
	& \multicolumn{3}{c}{nvGRAPH}& \multicolumn{3}{c}{Sphynx improvement}\\ \cmidrule(lr){2-4} \cmidrule(lr){5-7}
	&time	&cut	&imb	&speedup	&cut	&imb \\  \midrule
ecology1	&3.08	&45.6K	&1.82	&2.83x	&1.83x	&1.82x\\ 
dielFilter	&4.83	&1.9M	&1.89	&1.94x	&0.87x	&1.85x\\ 
thermal2	&3.69	&74.5K	&2.75	&2.91x	&1.66x	&2.75x\\ 
Bump\_2911	&9.13	&5.2M	&1.87	&2.28x	&0.90x	&1.85x\\ 
Queen\_4147	&9.06	&14.5M	&1.67	&1.34x	&1.00x	&1.65x\\ 
100$^3$	&2.98	&1.0M	&1.53	&1.92x	&0.80x	&1.51x\\ 
200$^3$	&17.87	&4.2M	&1.75	&2.75x	&0.82x	&1.73x\\ 
\textbf{geomean} & & & & \textbf{2.21x} & \textbf{1.07x} & \textbf{1.85x} \\
	    \bottomrule
    \end{tabular}
 }
  \end{center}
  \label{t:nvgraph}
  \vspace{-1ex}
\end{table}

We ran the above-mentioned experiments on all test graphs, however, nvGRAPH execution did not successfully finish on the large graphs while Sphynx ran successfully on all of them. 
Another issue is that nvGRAPH's imbalance is quite large on irregular graphs, where the maximum part weight in the resulting partition is at least about 20x larger than the average part weight.
We believe this is due to nvGRAPH minimizing a different metric, either the ratio cut or the normalized cut \cite{Naumov16}. These cut metrics trade off edge cuts and balance, and do not impose any strict balancing constraint during the partitioning.
Therefore, a direct comparison between Sphynx and nvGRAPH is difficult. Table~\ref{t:nvgraph} only displays the results on regular graphs, where imbalance results of nvGRAPH are still reasonable enough. We include the table as nvGRAPH is widely available, and the algorithm used is similar to ours.

As seen in Table~\ref{t:nvgraph}, Sphynx is 2.21x faster than nvGRAPH and obtains 1.07x smaller cutsize on average.
The average improvement obtained by Spynx on imbalance is 1.85x. 
The imbalance values obtained by nvGRAPH are ranging between 1.53 to 2.75, while the maximum imbalance value obtained by Sphynx is only 1.02.
Note that having a looser imbalance constraint enlarges the solution space so the partitioning tool is more likely to find a smaller cutsize.
This is probably why nvGRAPH finds a smaller cutsize on four graphs, yet, Sphynx still achieves to obtain a smaller cutsize on average.

%% file: conclusions.tex
\section{Conclusions and Future Work}
\label{sec:conc}

We have presented a new parallel spectral graph partitioner called Sphynx, which
runs on GPUs on distributed-memory systems. We evaluated its performance with 
Jacobi, polynomial, and algebraic multigrid (MueLu) preconditioners on both regular and irregular graphs,
and optimized it by exploiting several algorithmic choices. 
We observed that the combinatorial eigenvalue problem performs better on regular graphs
with all tested preconditioners. On irregular graphs, the generalized problem performs better
with Jacobi and MueLu, whereas the normalized problem performs better with polynomial preconditioning.
We observed MueLu was a highly effective preconditioner for regular graphs and polynomial was very effective for irregular graphs.
On regular graphs, setting the tolerance to 1e-3 with Jacobi and polynomial preconditioners controls the quality-vs-time tradeoff well, whereas in all other configurations, setting the tolerance to 1e-2 suffices.
The experimental results on Summit show that Sphynx is the fastest alternative on irregular graphs in an application-friendly setting and obtains a partitioning quality close to ParMETIS on regular graphs.
When compared to nvGRAPH on a single GPU, Sphynx is faster and obtains better quality partitions.  

An advantage of using a preconditioned eigensolver is we can use any
preconditioner for linear systems; therefore, we can easily benefit from
new developments in linear solvers and preconditioners.
We compared three important preconditioners but other methods such as domain decomposition and incomplete factorizations may also be used. Note that
since the matrices of interest are graph Laplacians, special preconditioners
with provably good convergence have been developed for this case \cite{BGHNT06,SpielmanTeng14}.
We leave this topic as future work, as such comparisons have already
been done in the context of solving linear systems \cite{BomanDeweeseGilbert16}.

Future work could include a study of time versus quality trade-offs for
spectral bisection and octasection, which would require more data movement
but might lead to better quality partitions.
The cut quality could also be improved by adding a refinement step,
for example greedy improvement. (Advanced refinement such as KL/FM 
is hard to parallelize, especially on GPU.)
Sphynx may also be used as a coarse partitioner within
a multilevel framework. This approach should be compared to Sphynx with
MueLu preconditioning, which can be viewed as a multilevel partitioning method.